\documentclass{PoS}
\usepackage{amsmath,graphicx}
\newcommand{\om}{\omega}
\newcommand{\Grec}{G_{\text{rec}}}

\newcommand{\ctune}{\cite{Morrin:2006tf}}
\newcommand{\cquench}{\cite{Umeda:2002vr,Asakawa:2003re,Datta:2003ww,Jakovac:2006sf}}

\title{Charmonium properties in the quark-gluon plasma}

\ShortTitle{Charmonium in the QGP}

\author{M.B.~Oktay, M.J.~Peardon, \speaker{J.I.~Skullerud}%
         \thanks{Address after 1 Sep 2007: Department of Mathematical
         Physics, NUI Maynooth, Co Kildare, Ireland.}\\
	 School of Mathematics, Trinity College, Dublin 2, Ireland\\
        E-mail: \email{jonivar@skullerud.name}}

\author{G.~Aarts, C.R.~Allton\\
        Department of Physics, Swansea University, Singleton Park,
	Swansea SA2 8PP, Wales, UK
}

\abstract{We present results for charmonium correlators and spectral
          functions in 2-flavour QCD on anisotropic lattices. Our
          results indicate
          that the S-waves ($J/\psi$ and $\eta_c$) survive up to temperatures
          close to $2T_c$, while the P-waves ($\chi_{c0}$ and
          $\chi_{c1}$) melt away below $1.2T_c$.}
\FullConference{The XXV International Symposium on Lattice Field Theory\\
                 July 30 - August 4 2007\\
                 Regensburg, Germany}

\begin{document}

\section{Introduction}

The properties of charmonium in the quark--gluon plasma is an issue of
long-standing interest, following the suggestion 
\cite{Matsui:1986dk} that $J/\psi$ suppression could be a probe of 
deconfinement. % Potential model calculations using the heavy quark free 
%energy have tended to support this picture.  
Lattice 
simulations in the quenched approximation \cquench\ have cast doubt
over this picture, indicating that $J/\psi$ may survive up to
temperatures as high as $1.5-2T_c$.
Recently, potential model calculations using the internal energy of the 
heavy-quark pair have reached the same conclusion, and indicate a
qualitatively similar picture in the case of $N_f=2$ QCD
\cite{Wong:2005be,Alberico:2006vw}.  This conclusion
has, however, been disputed by other potential model calculations
\cite{Mocsy:2007yj}.
%Support has also been provided by studies employing a real-time 
%static potential \cite{Laine:2006ns,Laine:2007gj} and a T-matrix approach 
%which includes scattering states \cite{Cabrera:2006wh}.

On the experimental side, data from SPS and RHIC show
similar amounts of $J/\psi$ suppression
at both experiments, despite the big difference in energy density.
Two different scenarios have been developed to explain this result.
The sequential suppression scenario \cite{Karsch:2005nk} takes its cue
from lattice results, suggesting that the entire observed suppression
originates from feed-down from the excited 1P and 2S states, which
melt shortly above $T_c$, while the 1S state survives up to energy
densities higher than those reached in current experiments.  On the
other hand, the regeneration scenario
\cite{BraunMunzinger:2000px,Thews:2000rj,Grandchamp:2002wp}
suggests that additional charmonium
is produced at RHIC energies from coalescence of $c$ and $\bar{c}$
quarks originating from different pairs.

Lattice simulations with dynamical fermions (2 or 2+1 light flavours) will 
be one of the essential ingredients in resolving several of these issues.  
In order to obtain a sufficient number of points in the euclidean time
direction with moderate computational resources it is necessary to use
anisotropic lattices, where the temporal lattice
spacing $a_\tau$ is much smaller than the spatial lattice spacing
$a_s$.  Here we present results from $N_f=2$ simulations with
anisotropy $\xi=a_s/a_\tau=6$.

\section{Methods}

We have carried out simulations on lattices with a temporal lattice
spacing $a_\tau\approx(7\text{GeV})^{-1}$ and an anisotropy
$\xi=a_s/a_\tau=6$, giving a spatial lattice spacing of $a_s=0.167$fm.
Configurations have been generated at a range of temperatures, from
220 GeV to 440 GeV, and two spatial volumes, $8^3 (1.34\text{fm})^3$
and $12^3 (2\text{fm})^3$.  500--1000 configurations were generated
for most temperatures.  The pseudocritical temperature for these
lattices was estimated to be $T_c=$205--210 MeV.  For details about
the anisotropic lattice action, see \ctune, and for further details
about the high-temperature configurations, see \cite{Aarts:2007pk}.

The spectral functions
$\rho_\Gamma(\omega)$ are related to the imaginary-time
correlator $G_\Gamma(\tau)$ (we consider here only zero momentum) according to
\begin{equation}
G_\Gamma(\tau) = \int_0^\infty \frac{d\omega}{2\pi}
 K(\tau,\omega) \rho_\Gamma(\omega),\qquad
K(\tau,\omega) = \frac{\cosh[\omega(\tau-1/2T)]}{\sinh(\omega/2T)}.
\label{eq:spectral}
\end{equation}
where the subscript $\Gamma$ corresponds to the different quantum
numbers.  Correlators were computed in the pseudoscalar, vector,
axial-vector and scalar channels, using local operators only, ie.\
$\Gamma=\{\gamma_5,\gamma_i,\gamma_5\gamma_i,1\}$.  Spectral functions
were then determined using the maximum entropy method (MEM)
\cite{Bryan:1990,Asakawa:2000tr}, with the modified kernel introduced in
\cite{Aarts:2007wj}, which greatly improved the stability and
convergence of the MEM analysis.

In addition to the spectral functions, we have also studied
reconstructed correlators $\Grec(\tau;T,T_0)$, obtained by integrating
(\ref{eq:spectral}) at temperature $T$ using the spectral function
$\rho(\omega;T_0)$ obtained at some reference temperature $T_0$.  If
the spectral function at $T_0$ can be obtained with a high degree of
confidence, this may provide a more robust way of analysing possible
medium modifications than by attempting to extract $\rho(\om)$ using
MEM at higher temperatures, where the number of data points decreases
as $1/T$.

\section{Results}

\begin{figure}
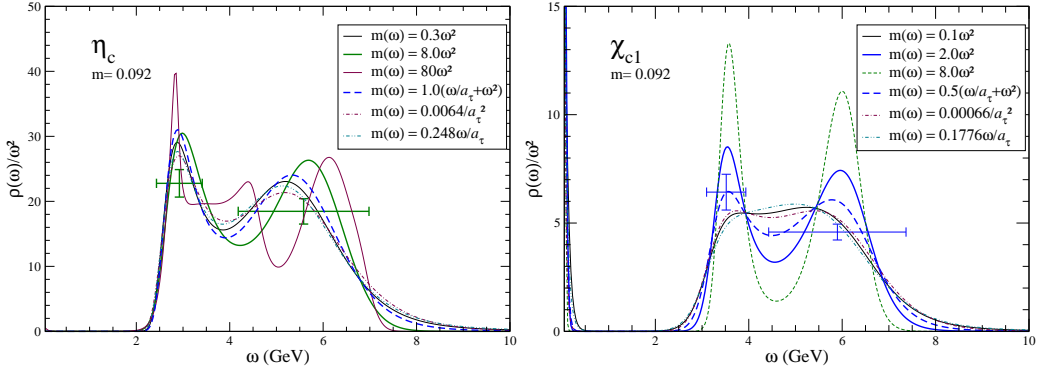

\includegraphics[width=0.45\textwidth]{PS_H_8x32_mw.eps}
\includegraphics[width=0.45\textwidth]{AX_H_12x32_mw.eps}
\caption{The pseudoscalar (left) and axial-vector (right) spectral
  functions for different default models, at $T=221$ MeV.}
\label{fig:baseline}
\end{figure}

The pseudoscalar and axial-vector spectral functions at $T=221$ MeV,
just above the deconfinement transition,
are shown in figure~\ref{fig:baseline}, for a range of different
default models.  In the pseudoscalar channel, we can see very little
dependence on the default model: only for the default model
$m(\om)=80\om^2$ is there any significant deviation.  Note that in the
graph, this model would corresponds to a horizontal line at 80, and in
the absence of information from the correlators the MEM would
reproduce this default model.  It is natural that the tension between
this rather extreme model and the data will produce some distortion in
the reproduced spectral function.  We conclude from this that the
pseudoscalar spectral function can be reliably reproduced at this
temperature.  In particular, we note that the position of the first
peak corresponds very closely to the zero-temperature $\eta_c$ mass
(the charm quark mass used in this analysis was somewhat lower than
the physical charm quark mass).  The second peak is assumed to be a
lattice artefact as its position is close to a sharp cusp in the free-field
spectral function.

In the axial-vector channel, on the other hand, we find a great deal
more variation.  Our `best' estimates for the spectral function (thick
lines) indicate a surviving peak structure approximately at the mass
of the zero-temperature $\chi_{c1}$, but the strong default
model dependence, combined with the proximity of the lattice artefact
peak, makes it impossible to distinguish with any confidence between a
bound state peak and a two-particle threshold.

As discussed above, we can be quite confident about our pseudoscalar
spectral function at $T=221$ MeV, and will therefore use this as a
reference temperature for studying reconstructed correlators.  The
same is the case in the vector channel, while in the P-wave (scalar
and axial-vector) channels the spectral functions at $T=221$ MeV are
much more uncertain.  We will nonetheless use this as our reference
temperature also in this case.

\begin{figure}
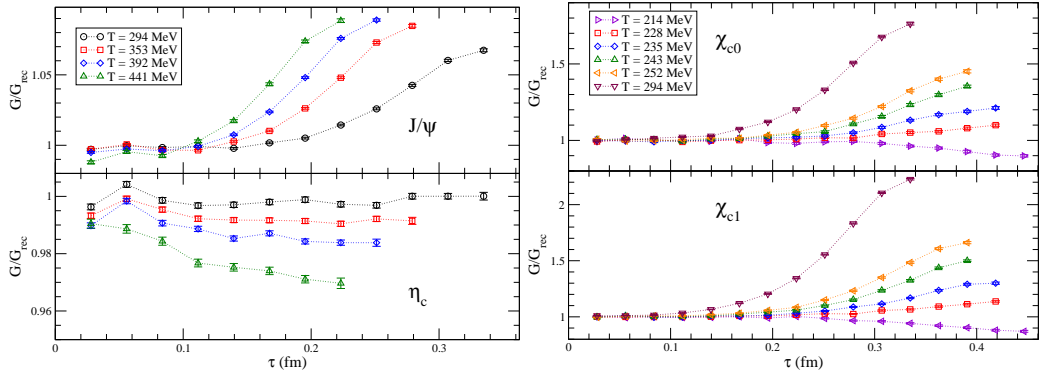

\includegraphics[width=0.45\textwidth]{Swave_8xNt.eps}
\includegraphics[width=0.45\textwidth]{Pwave_12xNt.eps}
\caption{Reconstructed correlators for the S-wave (left) and
  P-wave(right) channels.}
\label{fig:Grec}
\end{figure}

The reconstructed correlators are shown in figure~\ref{fig:Grec}.  We
can see that in the pseudoscalar channel, $\Grec(\tau)$ changes by only
2\% as the temperature is doubled with respect to $T_0$.  In the
vector channel the change
is slightly larger, but still less than 10\%.  Although it is quite
possible that a null result for $\Grec(\tau)$ can go hand in hand with
dramatic changes in the spectral functions \cite{Mocsy:2007yj}, this
provides prima facie evidence that medium modifications are small in
these channels.

In the P-wave channels the picture is dramatically different: the
long-distance correlators change by a factor 1.5--2 over a much
smaller temperature range.  This would indicate strong medium
modifications, although a large part of the change is most likely due
to the constant mode which appears in these correlators above $T_c$
\cite{Umeda:2007hy}.   This will be investigated further in the
future.  It is worth noting that our results for the reconstructed
correlators are very similar to those obtained in the quenched
approximation for the same values of $T/T_c$.

\begin{figure}
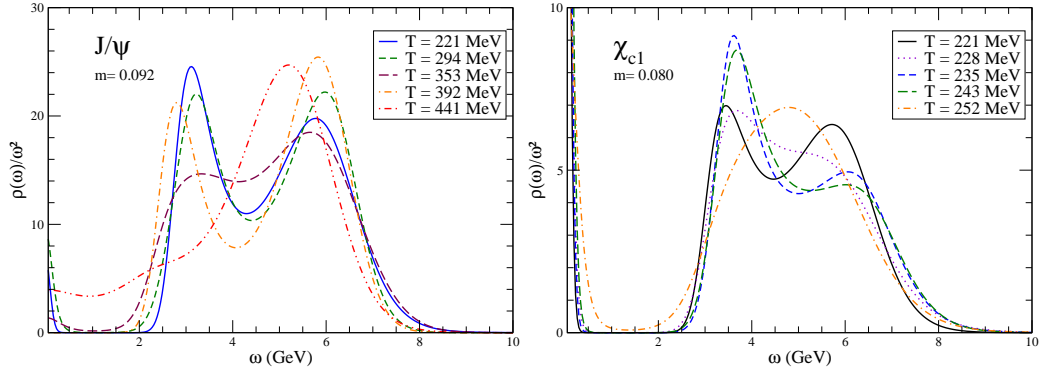

\includegraphics[width=0.45\textwidth]{Jpsi_R6m092.eps}
\includegraphics[width=0.45\textwidth]{chic1_R6m080.eps}
\caption{Spectral functions in the vector (left) and axial-vector
  (right) channel, for different temperatures.}
\label{fig:spectral}
\end{figure}

Figure~\ref{fig:spectral} shows spectral functions in the vector and
axial-vector channel at a range of different temperatures.  In the
vector channel, the ground state peak appears to survive unchanged up
to $T\approx300$ MeV, while at $T\approx440$ MeV there is no sign
of any surviving ground state.  The default model dependence
of our spectral functions becomes progressively stronger as the
temperature is increased, and as a result we are not in a position at
present to determine precisely where in the range $T=300-440$ MeV the
$J/\psi$ melts.

The nonzero intercept at $\om=0$ at our highest temperature could be
related to a non-vanishing charm diffusion coefficient.  We have
attempted to study this using the default model $m(\om)=m_0\om(b+\om)$
which explicitly allows $\lim_{\om\to0}\rho(\om)/\om\neq0$
\cite{Aarts:2007wj}.  However, our results at present are
inconclusive, and finer lattices will in all likelihood be required to
make any clear statements.

The spectral function in the pseudoscalar channel is
qualitatively similar to the vector channel, but the ground state
appears to survive to somewhat higher temperatures. 

In the axial-vector channel there is no sign of any surviving ground
state peak for $T\gtrsim250$ MeV or $1.2T_c$.  Below this temperature
there appears to be a peak structure which may correspond to a
surviving $\chi_{c1}$ state or to a threshold enhancement as discussed
above.  We conclude from this that the $\chi_{c1}$ melts below $250$
MeV, although a more careful analysis taking into account the constant
contribution in the correlator will be required to draw any firm
conclusion.  A further complication is that the local operator has a
poor overlap with the $\chi_{c1}$ state, making it more difficult to
extract information about the ground state if it exists.  This could
be improved on by using a derivative operator in the axial-vector as
well as in the scalar channel.
The picture in the scalar channel is similar to that in the
axial-vector channel, but the uncertainty is greater.

\section{Outlook}

Our results indicate that charmonium S-waves ($J/\psi, \eta_c$)
survive in the quark--gluon plasma up to 1.4--2 times the critical
temperature, while P-waves ($\chi_c$) melt below $1.2T_c$.  This is
conistent with what has previously been found in the quenched
approximation.  There are a number of uncertainties that must be
addressed before firm conclusions can be drawn.
\begin{itemize}
\item First and foremost, the spatial lattice spacing used in this
  first study is fairly large (0.17fm).  This means that the number of
  temporal points in the relevant temperature range is quite small,
  even with an anisotropy of 6,
  limiting the usefulness and reliability of the maximum entropy
  method.  Also due to the coarse lattice, lattice artefacts appear in
  the spectral function uncomfortably close to the physical ground
  state masses, obscuring the physical interpretation of the spectral
  functions extracted using MEM.  Simulations with finer lattices are
  underway, which should go a long way towards addressing these
  issues.
\item The simulation has used two degenerate flavours of sea quarks
  with a mass around the strange quark mass, rather than two light
  quarks and one strange quark as in nature.  The main effects of this are
  to raise the (pseudo-)critical temperature above its physical value,
  and to inhibit charmonium dissociation by light quarks.  Simulations
  with lighter sea quarks are planned.
\item The P-waves should be probed with operators which have a better
  overlap with the states of interest.  We are currently producing
  P-wave correlators using derivative operators, which we expect will
  provide more robust results for these channels.
\end{itemize}

We are also computing charmonium correlators at nonzero momentum.
This will allow a more direct comparison with experimental results,
where all $J/\psi$ particles that are observed (as dileptons) in the
detectors have nonzero (transverse) momentum.  Most importantly, it
may inform the interpretation of the data in terms of models such as
the sequential suppression model and the regeneration model, since the
momentum and rapidity dependence of the $J/\psi$ yields is a key
factor in distinguishing between these different models.

\section*{Acknowledgments}

 This work was supported by the IITAC project, funded by the Irish
Higher Education Authority under PRTLI cycle 3 of the National
Development Plan and funded by IRCSET award SC/03/393Y, SFI grants
04/BRG/P0266 and 04/BRG/P0275.  G.A. was supported by a PPARC advanced
fellowship.  We are grateful to the Trinity Centre
for High-Performance Computing for their support.

\bibliography{trinlat_bib/trinlat,spectral}

%\begin{thebibliography}{99}
%  \bibitem{...} ....
%\end{thebibliography}

\end{document}